\documentclass[preprint]{aastex}
\newcommand{\Sub}[1]{_\mathrm{#1}}

\newcommand{\ie}{{\it i.e.\@ }}
\newcommand{\rmm}[1]{\mathrm{#1}}

\shorttitle{The interaction of an AGB wind and a Jet.}
\shortauthors{Garc\'{\i}a-Arredondo \& Frank}

\begin{document}

\title{Collimated Outflow Formation via Binary Stars. \\
  3-D Simulations of AGB Wind and Disk Wind Interactions}
\author{F. Garc\'{\i}a-Arredondo \and Adam Frank}
\affil{Department of Physics and Astronomy, University of Rochester}
\affil{Rochester, NY 14627-0171}
\email{fgarcia@pas.rochester.edu, afrank@pas.rochester.edu}

\begin{abstract}
  We present three-dimensional hydrodynamic simulations of
  the interaction of a slow wind from an asymptotic giant branch(AGB)
  star and a jet blown by an orbiting companion. The jet or
  "Collimated Fast Wind" is assumed to originate from an accretion
  disk which forms via Bondi accretion of the AGB wind or Roche lobe
  overflow. We  present two distinct regimes in the wind-jet
  interaction determined by the ratio of the AGB wind to jet
  momentum flux. Our results show that when the wind momentum flux
  overwhelms the flux in the jet a more dis-ordered outflow outflow
  results with the jet assuming a corkscrew pattern and multiple
  shock structures driven into the AGB wind.  In the opposite regime
  the jet dominates and will drive a
  highly collimated narrow waisted outflow.  We compare our results
  with scenarios described \citet{SOR00} and extrapolate
  the structures observed in PNe and Symbiotic stars.
\end{abstract}


\maketitle
\section{Introduction}\label{sec:intro}

The role of binary stars in the shaping of planetary and
proto-planetary nebulae (PN, pPNe) has been a topic of debate for
many years \citep{BGM00}. PNe occur in a wide variety of shapes
ranging from spherical, elliptical, narrow-waisted bipolar and
bi-lobed (an elliptical core with projecting bipolar lobes). Some
PNe also show narrow jet-like features. pPNe show striking bipolar
or jet like morphologies and recently this brief evolutionary
phase has been identified as the period during which much of the
shaping occurs. Much of the theorizing on how pPNe and PNe obtain
their shapes has included, either implicitly or explicitly, the
presence of a binary companion (see \citet{BaF02} for a review of
PNe shapes and shaping mechanisms).

Hydrodynamic Generalized Interacting Stellar Wind (GISW) models,
in which successive wind phases from a single evolving AGB star
sculpt the nebula, were believed to hold considerable promise
for some time \citep{Bal87,IPB89}. In these
scenarios fast winds from the PNe central star expand into an
aspherical slow wind expelled by the AGB progenitor. These models did
not explicitly require the presence of a companion but it was often
assumed that the shaping of the slow wind occurred via tidal or common
envelope interactions \citep{Liv93}.

While hydrodynamic models were successful at recovering many
elliptical and bi-lobed nebula, simulations demonstrated that the
more extreme "wasp-waisted" bipolars could not be easily recovered
\citep{IBF92,MEL95,DCB96}. In addition many PNe and pPNe
(particularly, but not limited to, those with jets) show
point-symmetric morphologies in which nebular features are
reflected around the central star \citep{SAT98}. Such features are
difficult to explain with purely hydrodynamic models. More
recently a new problem has emerged as it has been recognized that
radiation wind driving alone can not be responsible for the
momentum and energy budgets associated with pPNe \citep{BCA01}.

The failure of pure hydrodynamic models have led researchers to
consider MHD scenarios for PNe formation.  Of particular interest
are models in which the magneto-centrifugal forces from an
accretion disk \citep{KPU00} both launch the wind and collimate it
into a jet. \citet{BFW01} and \citet{FPG03} have shown that MHD
disk wind models can be quite effective at driving outflows with
the total momentum and energy observed in pPN and PN. We note that
weak magnetic field models are, by their nature, unable to produce
observed pPN momenta and we do not consider them here \citep{GLR99}. 

Of course, any model which requires an accretion disk will, in the
context of PNe, require a binary system. Observationally there
exist strong links which argue that disks winds and jets from binary
companions play a significant role in pPNe and PNe shaping. Close
binary PNe central stars are know to exist \citep{BON00} indicating
that interactions between stellar mass loss processes are
likely to occur at some point. More telling are the example of the
symbiotic stars.  These are binary stars in which a compact companion is
known to orbit an AGB star. There a numerous examples of highly
collimated outflows from these systems that appear quite similar to
PNe \citep{COR00}.

\citet{MOR87} and \citet{SOL94} where the first to identify
accretion disks from binaries as the source of highly collimated
PNe. A number of studies have explored the formation and
properties of the these disks \citep{MAM98,MAM99} and the nature of
the binary systems which would form them.  More
recently \citet{SOR00} have considered the existence of
Collimated Fast Winds (CFWs) from an orbiting companion's disk and
sketched out the flow pattern which would occur as the CFW
interacts with the AGB wind. \citet{LSO01} applied such a
model to M2-9 demonstrating that the resulting flow pattern could,
in principle, be embraced by the interaction of a CFW and AGB
wind.

In this paper we use 3-D adaptive mesh hydrodynamical simulations
to explore the hydrodynamics of AGB wind/CFW interactions. In
particular we focus on explicit realizations of the flow patterns
near the two stars in an attempt to understand the limits of this
class of models. Carrying forward simulations such as these pose a
number of numerical challenges and here we
attempt to explore only the fundamentals of the hydrodynamics. We note that
our models represent the first numerical exploration
of this important class of model and should serve to help
articulate key issues related to their application and further study.

The structure of our paper is as follows.  In section II
we describe the numerical model, assumptions and simplifications used in our
calculations.  In section III we describe our results and in section IV we
present a discussion of their implications as well as our conclusions.

\section{Numerical Simulations}
\label{sec:numet}
\subsection{Numerical methods and equations}
\label{subsec:numeq} We work in the regime of ideal gas dynamics
in the presence of gravitational sources.  Thus we solve the Euler
equations, which, in conservative form are
\begin{equation}
  \frac{\partial \rho}{\partial t} + \nabla \cdot (\rho
       {\bf v}) = 0 \hskip .3truecm ,
       \label{eq:1}
\end{equation}

\begin{equation}
  \frac{\partial \rho {\bf v}}{\partial t}+ \nabla
  \cdot (\rho {\bf v v} + p{\bf I}) = -\rho \nabla \Phi \hskip
  .3truecm ,
  \label{eq:2}
\end{equation}

\begin{equation}
  \frac{\partial E}{\partial t} + \nabla \cdot ({\bf
    v}[E+p]) = -\rho {\bf v} \cdot \nabla \Phi \hskip .3truecm ,
  \label{eq:3}
\end{equation}
where $\rho$, ${\bf v}$ and $p$ are the density, velocity and
pressure at a point of the fluid, respectively, ${\bf v v}$ is the
tensor product and ${\bf I}$ is the unit tensor. $E$ is the total
energy per unit of volume, given by
\begin{equation}
  E=\frac{p}{\gamma - 1} + \frac{1}{2} \rho |{\bf v}|^2
  \hskip .3truecm .
  \label{eq:4}
\end{equation}
{In addition, we had included an advection equation for a passive
scalar, which will be used for tracing the different flows. Here,
gravity is due solely to the stellar sources and self gravity is
not included. The gravitational source term is given by

\begin{equation}
  \nabla \Phi = G \left(\frac{M\Sub{AGB}{\bf r}}{|{\bf
      r}|^3} + \frac{M\Sub{c} ({\bf r} - {\bf r\Sub{c}})}{|{\bf r} -
    {\bf r\Sub{c}}|^3}\right) \hskip .3truecm ,
  \label{eq:5}
\end{equation}
where $M\Sub{AGB}$ and $M\Sub{c}$ are the masses of the AGB and its companion,
respectively, and ${\bf r\Sub{c}}$ is the position of the companion.
The gravitational source term is handled in an operators split
fashion via a first order integration carried out between
hydrodynamic time-steps.

The numerical simulations are carried out using a version of the
yguaz\'u-a adaptive code, described by \citet{Rag00}. This code is
well tested and has been used for a variety of supersonic flow
problems. The hydrodynamic equations are numerically integrated on
an adaptive mesh refined using a second order flux vector
splitting scheme with a \citet{Van82} algorithm. We used four
levels with $640 \times 320 \times 320$ cells on the maximum
level. The boundary conditions are transmission for both $x$ and
$y$ boundaries as well as the  maximum $z$ boundary. A reflecting
boundary condition is used on the lower $z$ axis, due to the
symmetry of the problem.

We have carried out our simulations assuming an approximate
isothermal flow. Thus we choose $\gamma = 1.01$. Since our
simulations are carried out close to the stars the densities in
the flow are quite high ($\rho \approx 10^{-18} ~\rmm{g}$
$\rmm{cm}^{-3}$) and the cooling times will be shorter than
what are usually considered in nebular problems.  In spite of the
high densities the temperature structure behind high velocity shocks,
$V\Sub{s} \approx 1000 ~\rmm{km~s^{-1}}$, are likely to be complex and
post-shock gas may not immediately cool.  In such cases a
time-dependent treatment of the cooling should be used. In this paper
we only explore cases with lower velocities where the cooling
timescales will be relatively short compared with dynamical times and
the isothermal assumption is acceptable.

\subsection{Initial Conditions}
\label{subsec:inico} Our simulations were designed to explore the
interaction of a collimated wind from an orbiting companion
interacting with the spherical outflow from the AGB primary. We do
not attempt to explore the formation of accretion disks and simply
apply an inflow condition at the location at the instantaneous
position of the companion.  As we discuss below however, we have
chosen binary parameters which are expected to lead to the
formation of an accretion disk.  We note that these simulations
are computationally intensive. We made a number of assumptions and
simplifications which allowed a trade-off between remaining in the
correct parameter regime and resolution/computation resource
issues.

For the AGB wind we assume typical values of parameters. We begin
with a $1.4$ M$_\odot$, star driving a spherically symmetric,
mass-loss of rate $\dot{M}\Sub{AGB} = 10^{-6}$ M$_{\odot}$
yr$^{-1}$. We take the radius of AGB to be a "wind release" radius
of $r\Sub{AGB} = 4 $~AU.  This is the point at which we inject the
wind into the grid with $v\Sub{AGB}(r\Sub{AGB}) >
v\Sub{esc}(r\Sub{AGB})$. Given the computational requirements needed
for the hydrodynamics alone we do include radiation driving on dust in
the wind which would produce a gradual acceleration until an terminal
velocity is reached at $r>>r\Sub{AGB}$.  Instead the wind follows a
type III solution to the classic parker wind equation \citep{ARN98}
\begin{equation}
  (v^2-c^2)\frac{dv}{dr} = \left(\frac{2c^2}{r} -\frac{GM}{r^2}
  \right) v \hskip .3truecm ,
  \label{eq:solsta}
\end{equation}
where $c$ is the sound velocity.  Given our boundary conditions at
$r\Sub{AGB}$ our wind first decelerates due to gravity and then
reaccelerates on a trajectory similar to the radiation-driven
wind. Thus the radial wind structure of the wind close to the star
departs from dust-driven wind models. Note that the wind is always
supersonic. We have carried out simulations in which we move the
inflow boundary further outward such that the wind velocity
remains relatively constant and have found no change in the global
flow patterns.

For our companion we chose a $.6$ M$_\odot$ star with orbital
separation $a = 10$ AU.  We assumed a spherical orbit with
resulting orbital period of T$\Sub{o}\approx 22.4$ years. We note
that our simulations are carried out with the AGB primary fixed to
the origin of the coordinate system.  We do not attempt to treat
the reflex motions of the companion. This greatly reduces the
complexity of the boundary conditions on both AGB star and the
jet. Comparison of the various velocities in the problem
demonstrates that such an assumption should not change the global
flow patterns we attempt to explore. If
$v\Sub{1}=\sqrt{G}M\Sub{c}(M\Sub{AGB}+M\Sub{c})^{-1/2}a^{-1/2}$ is
the velocity of the primary about the center of mass we require a
hierarchy of velocities $v\Sub{j} >> v\Sub{AGB}>v\Sub{1}$.  For
our parameters we find that at $v\Sub{1}/v\Sub{AGB}=0.16$.
Therefore, we expect that the reflex motions should not have a
dramatic effect our results.

Based on the description of \citet{SOR00} we have
chosen to explore two cases for the properties of the jet. For
"Weak Jet" case, the momentum flux from the AGB star is larger
than the that from the jet. We define a parameter $\chi$

\begin{equation}
  \chi = \frac{\dot{M}\Sub{AGB} v\Sub{AGB}}{\dot{M}\Sub{j} v\Sub{j}}
  \label{momrat}
\end{equation}

\noindent where the subscript "j" refers to properties of the jet.
The Weak Jet models have $\chi > 1$ whereas the "Strong Jet" case
has $\chi < 1 $. In the two Weak Jet simulation we present we used
$ v\Sub{j}= 200$ km $\rmm{s}^{-1}$, $\dot{M}\Sub{j}=10^{-7}$
M$_{\odot}$ yr$^{-1}$, and $\dot{M}\Sub{j}=2.5\times 10^{-8}$
M$_{\odot}$ yr$^{-1}$, which yield $\chi=1.25$, and $\chi= 5$.  In
the Strong Jet case the $ v\Sub{j}= 400$ km~s$^{-1}$, and
$\dot{M}\Sub{j}=10^{-7}$ M$_{\odot}$ yr$^{-1}$ yielding
$\chi=0.625$. We note that values for the parameters taken for the
jet are appropriate to an outflow driven by a main-sequence
companion where $v\Sub{j} \approx v\Sub{esc}$. In our simulations
the jet is not fully collimated. At the instantaneous position of
the companion we inject a flow with the angle of collimation 10
degrees. Note also that our strong jet parameters imply that the
companion accretes a large fraction of the AGB wind.  This may not
be realistic but was chosen to allow us to use a lower jet speed
and maintain the isothermal assumption for the computations.

To have a produce a collimated jet it is necessary for an
accretion disk to form around the secondary star. As discussed in
\citet{SOR00} and elsewhere, a disk can be formed when the
specific angular momentum of accreted material $j\Sub{a}$ is
larger than the specific angular momentum of the accretor $j_2$
(the companion in a Keplerian orbit) \ie $j\Sub{a}/j_2>1$.  From
\citet{SOR00}

\begin{eqnarray}
  \frac{j_a}{j_2}=15\left(\frac{\eta}{0.2}\right)
  \left(\frac{M\Sub{AGB}+M\Sub{c}}{1.2 M_{\odot}}\right)^{1/2} \nonumber \\
  \times \left(\frac{M_2}{0.6 M_{\odot}}\right)^{3/2}
  \left(\frac{R_2}{0.01R_{\odot}}\right)^{-1/2} \nonumber \\
  \times \left(\frac{a}{10~\rmm{AU}}\right)^{-3/2}
  \left(\frac{v\Sub{r}}{15~\rmm{km~s^{-1}}}\right)^{-4}
\end{eqnarray}

where $\eta$ is the ratio of the specific angular momentum of
accreted material to that in material that enters the Bondi-Hoyle,
cylinder \citep{LIV86,SOR00}. $v\Sub{r}^2 = (v\Sub{AGB}^2 +
v_2^2)$ is the relative velocity of the wind and companion.  We
take $\eta = 0.2$ which applies to a case where the accretion lies
between an fully isothermal and adiabatic flows and $R_2 =
.6~R_{\odot}$.  At $r = a$ we find a wind speed of
$v\Sub{r}\approx 15~\rmm{km~s^{-1}}$  Thus we find
$j\Sub{a}/j_2\sim 2.5$ implying that a disk would form via
accretion of the AGB wind.  Detailed SPH simulations carried out
by \citet{MAM98} using similar parameters as ours tracked the
accretion flow and demonstrated that the formation of accretion
disks was a robust result. An upper limit to the size of the disk
can be calculated using the Bondi-Hoyle radius $R_a =
2GM\Sub{c}/v\Sub{r}^2 = 7\times10^{13}~\rmm{cm}$.  A more exact
relation for the disk radius is given in \citet{LSO01}, from which
we find $R\Sub{d}\approx 1.278 \times 10^{12}$ cm. This size of
the disk is less than 1 pixel in our simulation ($\Delta x \approx
.5 ~\rmm{AU}$ at the highest resolution).

\section{Results}
\label{sec:res} In this section explore the flow pattern obtained
in the Weak and Strong Jet cases and attempt to link these with
previous theoretical and observational works.

\subsection{Strong Jet}
\label{subsec:strongjet} We begin with consideration of the strong
jet case whose flow pattern is simpler to visualize and
understand. In Figure~\ref{fig:fg1} we present density
grey scale maps of the Strong Jet simulation in the: (a)
$xz$-plane: (b) $yz$-plane: (c) $xy$ plane. The first two cross-
sections run through the central axis of the computational space
and the third is taken at the base of the flow and includes the
AGB and companion/jet inflow boundaries.
Figure~\ref{fig:fg1} shows this simulation after 224
years, or 10 orbital periods.

With a momentum flux ratio of $\chi = 0.625$ one would expect that
the jet material would be able to propagate through the AGB wind
relatively unimpeded.  If shocks do occur they will be found along
the sides of the expanding jet column where the jet has pushed AGB
material away, entraining and accelerating it. Most importantly we
would do not expect the jet to be strongly deflected. The
principle modification to jets motion could arise from the shift
in its launch point due to the orbital motion of the companion.
This will only occur when the orbital period $T_o$ is comparable
to the jet crossing time though the computational domain
$T\Sub{cross} =H/v\Sub{j}$ where $H$ is the height of the
computational space.  When $T\Sub{cross}/T\Sub{o} \ge 1$ then a
corkscrew morphological pattern may be expected for the jet's 3-D
structure. The spherical AGB wind, on the other hand, will be
strongly modified by the passage of jet material. AGB material
whose trajectories cross the jet will have to be shocked and
diverted.  Since we do not expect strong shocks in the jet beam
for our Strong Jet case,  the velocity $v\Sub{j} = 400
~\rmm{km~s^{-1}}$ is expected all the way up the jet. Thus
$T\Sub{cross}/T\Sub{o} = .1$ and we will not expect much
"twisting" in the jet morphology.

These expectations are born out by the simulations. Consider the
$xz$ and $yz$ cross section in Fig~\ref{fig:fg1} (panel a and b).
Panel a) shows a broad ${\bf V}$-shaped outflow "lobe" driven by
the jet as it pushes through the AGB wind. The outer boundaries of
the ${\bf V}$ are shock waves propagating through the expanding
AGB wind. This shock creates a dense shell of compressed AGB
material (see left panel detail). Note that this structure is not
symmetric. The concave shell appears as a darker "bar" on the
right side on plot a) with a weaker feature on the left. Note that
similar pattern appears in plot b) but here the jet only appears
at a higher value of $z$ and there are stronger asymmetries
between the left and right part of the grid. The asymmetry is a
result of the 3-D nature of flow.  Shocks are driven through the
AGB as the jet sweeps through its orbit. Thus a given point in the
computational space experiences the passage of a succession of
strong pressure wave at intervals of approximately $T\Sub{o}$.
Note also the high density structure which can be traced back to
the jet source with a second low density region external to it
(the lower-left panel of Figure~\ref{fig:fg1}. This feature may be
due to the movement of the jet inflow boundary condition as new
cells become injection points while others return to being
reflection boundaries.

Another, nested, ${\bf V}$-shaped feature comprised of lighter
greyscales is also apparent. This defines the limits of the cavity
carved out by the jet. Consideration of the details of the flow
patterns show that propagation of the jet through the AGB material
results in considerable entrainment of AGB gas (the darker
greyscales immediately surrounding the jet) and the final state is
one in which the accelerated AGB material fills a large volume of
the cone defined by the global flow.  This entrainment appears to
occur via the oblique shocks which define interface of the jet/AGB
flow.  As the jet material is slowed and redirected via these
shocks AGB material is accelerated and mixed into the shell which
defines the cavity.  Note the periodic structures at the edges of
the jet. We conjecture that these features represent
Kelvin-Helmholtz modes which are only just resolved in these
simulations.  These modes will enhance AGB acceleration and mixing
at the interface of jet/AGB cavity.

Examination of Figure~\ref{fig:fg1} shows how the orbital motion
of the jet drives a positional interchange between the current
location of the companion and the location where the strongest
jet-AGB wind interaction will occur. In the enlargement of the
lower region of panel (a), the $xy$ cross section, we see the jet
lies to the right of the AGB star. The lower density (lighter
greyscale) material can be seen flowing in the $z$-direction in
panel (a). Jet material which has been ejected earlier (and is at
higher $z$) will be further up the beam and deeper into the 3-D
volume then the position of this $xy$ cut. The dark greyscales
which bound the jet are "sideways" shocks which result when ever a
jet flow changes its directions. Such features have been seen and
described in simulations of precessing jets \citep{RCB93,CFJ96}.
The shift in position of the jet and shock features is also
apparent in panel (b) which shows the $yz$ plane. Here the lower
density features which appear above the AGB star is jet material.
This structure occurs both because the opening angle of the jet
and its orbital movement. In this simulation the jet is strong
enough that its core remains undisturbed until it leaves the grid.
We do some mild orbital effects in the jet morphology reflected in
the apparent change in direction of jet beam in panel a) at the
top of the grid where the lightest contours terminate.  In the top
of panel b) however the lightest contours representing undisturbed
jet extend to the edge of the grid.

The global morphology of the jet-AGB wind interaction can be seen
in Fig~\ref{fig:fg2}. This figure shows a 3-D volume rendering of
the velocity along with an iso-surface represented as a wire-frame
mesh. Here we can clearly see that the jet has only a small trace
of a corkscrew pattern and that it is the expansion of the jet and
the entrainment of AGB material which contribute most strongly to
the velocity (and hence density) pattern. The entrainment of
shocked AGB wind with shocked jet material is particularly
apparent in this figure as the jet is surrounded by lower velocity
gas. We see shocked/mixed AGB material at a wide range velocities
with the highest speeds reached being $ v \sim 250$ km s$^{-1}$.
Note also the arc of material defined by the wireframe. This
feature defines the wake of the jet and companion though the AGB
gas via spiral shocks.

The behavior of the flow near the orbital plane displays a number
of interesting features. Panel (c) which shows the base of the
computational plane reveals spiral shocks which occur due to the
gravitation effect of the companion. While our simulation can not
resolve the Bondi radius of the companion and so can not track
wind accretion we note that spiral features such as these were
also seen in SPH simulations of disk formation \citep{MAM98}. Note
that we see these spiral shocks extend above the equatorial plane in
Figure~\ref{fig:fg2} just as did \citet{MAM98}. In our case
however the spiral shocks above the plane appear closely connected
with the wake of the jet. We note that a ring or torus of higher
density compressed AGB material appears near the equator. This
material is formed via the action of the rotating jet (apparent in
Figures~\ref{fig:fg1} and~\ref{fig:fg2} and was
predicted by \citet{SOR00}).

In Fig~\ref{fig:fg3} we show 1-D cuts of density and velocity
through the computational space in the $xz$ plane at $x=80$ AU,
which corresponds to the center of the AGB star, and $x=90$ AU,
which corresponds to the initial symmetry axis of the jet. The
$x=80 ~\rmm{AU}$ cut shows a steep rise in velocity which begins
at $z=20~\rmm{AU}$ and terminates at $z=80~\rmm{AU}$. This feature
reflects the acceleration of the AGB material as it interacts with
the jet. For $z>80~\rmm{AU}$ the line cuts though jet material
traveling at close to its injection velocity.  The $x=90$ cut
shows a velocity which is approximately constant reflecting the
fact that the jet is strong enough that its core remains
undisturbed by the AGB material. Note the oscillations in density
and velocity which reflect the apparent KH modes along the jet/AGB
interface.

To summarize, the interaction of the jet with the slow wind
produces a complex morphology composed of multiple shock waves
driven into both the AGB wind and jet.  In spite of its complexity
a quasi-steady state patten does emerge in which the entire flow
is embedded in a ${\bf V}$-shaped shock propagating outward
through the AGB wind.  The shock produces a dense shell of
compressed AGB material. Within this shell is a cavity produced by
the propagating jet which does not assume a strong corkscrew flow
pattern because the jet transit time is shorter than the orbital
period, \ie jet material moves fast enough that it leaves the grid
in a time less than an orbital period $t < T\Sub{o}$. Significant
entrainment of AGB and jet material occurs due to shocks at the
jet/AGB interface. At the equator spiral shocks are driven into
the AGB wind due to the orbital motion of the companion and the
effect of the jet.

\subsection{Weak jet}
\label{subsec:wejet} We now consider the simulation results for
the two cases in which in the momentum in the jet is weaker than
that in the AGB wind.   As the AGB wind momentum flux becomes
successively larger than that in the jet we expect strong shocks
to form in the jet beam as well as jet deflections caused the ram
pressure of the AGB material . The first weak jet simulation we
present has $\chi = 1.25$ and density maps in different orthogonal
planes are shown in the Figure \ref{fig:fg4}. The second case has
$\chi = 5$ and is shown in Figure 6. Once again, both simulations
corresponds to age of 224 years, or 10 orbital periods.

When the flux of momentum of the AGB wind is slightly higher than
that in the jet, $\chi =1.25$, the ${\bf V}$-shaped shocks in the
AGB wind marking the boundary of the interaction region still
appear. But, as shown in Fig \ref{fig:fg4}, the effects
of the stronger AGB wind are apparent. First it is possible to see
a shallow inclination of the jet in panel a) (see left panel
detail). For this value of $\chi$ the deflection of the entire jet
does not occur.  Instead, we see a narrowing of the jet on the
side facing the AGB in the region at low $z$, close to the jet
source. In the language of \citet{SOR00} the deflection
angle of the jet is still smaller than the collimation angle for
these parameters.  Thus only the collimation on one side of the
jet is affected by the AGB ram pressure.

What is more apparent in this case, as compared to the Strong Jet
simulation, is the strong shock in the jet beam.  We see a shock
and dense shell in panel a) at $z=100~\rmm{AU}$. Note that the
apparent termination of the jet at $z= 100 ~\rmm{AU}$ in panel a),
and $z=280 ~\rmm{AU}$ in panel b) are evidence of its 3-D
structure. In panel a) for example, jet material which has been
ejected earlier (and is now higher at $z = 100 ~\rmm{AU}$) will be
further up the corkscrew and deeper into the 3-D volume then the
position of this $xy$ cut. The dynamically important connection is
that the shock decreases velocities in the material surrounding
the jet.  $T\Sub{cross}/T\Sub{o}$ becomes larger and the orbital
motion begins to exert a greater influence on the jet morphology
in terms of creating a global "corkscrew" pattern.

Near the equatorial plane and close to AGB star, the flow
structure is very similar to the strong jet case. Above the AGB
star there is a shock wave and a thin shell of compressed
material. This shock can be better appreciated in the 1-D plots in
the Figure~\ref{fig:fg5}. Here the shock is clearly seen at
$z\approx 20$ AU at the $z$-axis ($x=80$ AU) and in $yz$- plane
(left panels in the Figure~\ref{fig:fg5}). After this shell,
the AGB material is gradually accelerated via entrainment reaching
a velocity of $\approx 180 ~\rmm{km~s^{-1}}$.  Using this value in
the expression for the crossing time we find $T\Sub{cross}/T\Sub{o}
\approx 0.4$.

When $\chi$ increases the deflection angle should increase (see
\citet{SOR00}, though their expression for the jet   
bending only applies when the bend occurs on scales smaller than
$a$). Thus for the $\chi=5$ simulation we expect greater
influence of the AGB ram pressure on the jet beam. This
expectation is borne out in the $xz$ and $yz$ maps in Figure
\ref{fig:fg6} which clearly show that entire beam has bent away
from the AGB star. Globally the cross-sections show a succession
of staggered "donkey ear" shaped shocks and dense shells
propagating through the AGB wind. Such a pattern was predicted by
\citet{SOR00} and occurs due to the deflection (and disruption) of
the jet by the slow wind over an entire orbital cycle.

Consideration of the post-shock velocities in the jet and
surrounding material (Figure~\ref{fig:fg7}) shows that
$T\Sub{cross}/T\Sub{o} \approx 0.86$. With this in mind the enlargement of
the $xy$-plane map in Fig \ref{fig:fg6} allows us to see the
origins of the large scale flow pattern. The jet can be seen
emanating from the companion on the right side of the AGB star in
panel a). Unlike the Strong Jet case however the jet only
propagates to $z \approx 30$ AU before it is shocked via its
interaction with the AGB wind. The jet is bounded above by a dense
shell comprised of both shocked jet and shocked AGB material. The
continuation of this structure can be seen in the $yz$ cross
section on the left side of the AGB star.

On the $xz$ and $yz$ planes is possible see, three "ear shaped"
lobes, two on panel a) and one on panel b).  Each of these
features are part of a continuous 3-D structure.  What we see in
the cuts are cross-sections of shocks formed from the interaction
of the jet and AGB wind, each initiated at successive 1/4
$T\Sub{o}$ time delays . Note only three lobes are apparent in the
figure implying that jet material ejected in the first 1/4 of the
current orbital period has already left the grid. Unlike the
Strong Jet case, some of the jet material exits from the side of
the grid due to the deflection.

The three dimensional structure of the flow is shown in the Figure
\ref{fig:fg8} for the $\chi=5$ case. The iso-surfaces in density
show have values: $6.168 \times 10^{-18} \rmm{g~cm^{-3}}$, $5.607
\times 10^{-19} \rmm{g~cm^{-3}}$, and $1.78  \times
10^{-19}\rmm{g~cm^{-3}}$, respectively.  Lower density
iso-surfaces are located at higher $z$.  This figure demonstrates
the formation of global corkscrew flow pattern produced due to the
orbital motion of the source.  Note that we see less than a single
turn of the corkscrew as would be expected for the value of
$T\Sub{cross}/T\Sub{o}$.  The complexity of the flow is also
apparent in this figure. The action of multiple shocks passing
through the AGB wind and jet creates and a mix of structured and
disordered iso-density surfaces.  A "donkey-ear" shock is apparent
in the middle and lower regions of the computational space.  Note
also that the spiral shocks in the AGB wind and the global ${\bf
V}$-shaped (in cross-section) shock which bounds the jet/AGB wind
interaction is also apparent in this figure.  We note that
explorations of figures such as these show that the exact choice
of iso-density surfaces does not change the qualitative
conclusions that can drawn from this figure.

\section{Discussion \& Conclusions}
\label{sec:disc}
We have presented new simulations of the interaction of a slow AGB
wind with a collimated fast wind (CFW) driven by binary companion.
Our binary parameters were selected such that an accretion disk
would form even though we are unable to resolve such a flow
pattern.  We report three simulations segregated by the ratio of
the AGB wind momentum flux to the CFW momentum flux (denoted
$\chi$).  We simulated a Strong Jet case $\chi = 0.625$ and two
Weak Jet cases, $\chi = 1.25$ and $\chi = 5.0$.  The binary period
was $T_o = 22.4 ~\rmm{years}$ and we carried all simulations out for at
least $10 T\Sub{o}$.

We find $\chi$ to be an effective predictor for differentiating
the behavior of the simulations.  Strong jets are able to
propagate off grid ($Z\Sub{max} = 320 ~ \rmm{AU}$) without severe
deflections or disruptions. When $T\Sub{cross}/T\Sub{o} << 1$, as
it was in the Strong Jet case, the orbital motion does not effect
the global morphology of the jet. When the jet becomes
progressively weaker, and $\chi$ becomes larger, we see a trend
towards both stronger deflections, (the jet is bent away from the
binary), and stronger disruption, (strong shocks bounding the jet
beam). The location, in height, of the first shock moves closer to
the jet source as $\chi$ increases. Weaker jets also lead to more
complicated global flow patterns with features such as multiple
"donkey ear" shaped lobes appearing at well characterized
intervals in height.

Our simulations support the idea that collimated jets, formed
close to the central source, are the agents shaping some PN.  This
paradigm has steadily been gaining favor. \citet{MOR87} and
\citet{SOL94} both proposed that jets could be formed via
accretion disks in PNe. \citet{SAT98} proposed that jets formed
very early in the PNe or pPNe phase were responsible for the bulk
of the nebular shaping with the fast wind from the central star of
a PN merely burnishing the details of the nebular shapes. More
recently \citet{SOK02} and \citet{LES03} have both explored
hydrodynamic models of jets driving through circumstellar
environments. Our work is complementary to the \citet{LES03} study
in that we examine the flow pattern on smaller scales.
\citet{SOR00} investigated both the hydrodynamics and population
statistics of the CFWs and the outflows they would drive in binary
systems. They concluded that narrow waisted PNe are likely the
result of CFW/AGB wind interactions. \citet{LSO01} applied such a
model to M2-9 concluding that a weak jet with a severely deflected
wind was responsible for both the outflow shape and the shadowing
of ionizing radiation from the hot companion.

Our results are most directly relevant to the work of \citet{SOR00}
and can be seen as a test of the ideas put forth there.
We find that much of the flow pattern predicted in that work is
obtained in our simulations. We do see some evidence for the
creation of a region of enhanced density along the equator due to
the action of the CFW as predicted \citet{SOR00}. The
successive donkey ear lobes seen in the simulations were
anticipated by \citet{SOR00} as well. It is not yet
clear if their strong deflection flow pattern will be obtained in
real systems however since we are currently not able to resolve
details of the jet flow when shocks form very close to the inflow
source. We will take up this issue, along with the effects of wind
acceleration, in a future study.

Our most important conclusion however is that CFWs do appear as
good candidates for creating narrow waisted nebula. The Strong Jet
case, in particular, shows that the effect of a CFW is primarily
confined within a {\bf V} or {\bf U} shaped shock in the expanding
AGB wind. Such a jet will drive a very narrow waisted flow as
there is no expansion along the equator. Thus on large scales the
lobes driven by the jet will appear to pinch severely at the
outflow source.  We note that recent HST observations of the near
nuclear regions of M2-9 appear to confirm the flow patterns we see
in the Strong Jet case (Balick, private communication 2003) with
oppositely oriented {\bf V}-shaped features extending from the
unresolved central source out to scales of $10^{15} ~\rmm{cm}$.

This work constitutes a first attempt at mapping out the flow
dynamics of outflows driven via binary interactions.  Here we have
attempted to examine the global properties of the outflows on
scales ranging from $\approx 5$ to $300 ~\rmm{AU}$. While the use of
an AMR code allowed us to capture details such as shocks in the AGB
wind and jet beam, the resolution used and limited physics
included still leave many open questions.  These include the
nature of the flow when $\chi$ is very large, the role of
radiation driving of the AGB wind, the effect of ionizating
radiation and the effect of magnetic fields (which should be
present if the jets are magneto-centrifugally launched).  Inspite
of these limitations our simulations provide further support for
the argument that jets driven by accretion disks can explain many
features observed in bipolar PNe.\\

We thank Noam Soker for his helpful comments on a draft of this
work. B. Balick, P. Varniere, E. Blackman, H. van Horn and J.
Thomas for related discussions. Support from the Space Telescope
Science Institute, NSF grants AST-9702484, AST-0098442, NASA grant
NAG5-8428, DOE grant DE-FG02-00ER54600 and the Laboratory for
Laser Energetics. F.G.A. is grateful to CONACYT, M\'exico for a
posdoctoral scholarship.

\clearpage
\begin{figure}
  \caption{Strong Jet greyscale density maps of cross sections in
    the a) $xz$ plane, b) $yz$ plane, and c) $xy$ plane (\ie the
    equatorial plane). Panels on both sides of plot c) are enlargements of
    the plots a) and b). The white dots are the position of the center
    of the stars. The units in the axises are given in
    AU. The grayscale is the logarithm of the density.}
  \label{fig:fg1}
\end{figure}
\begin{figure}
  \caption{Three-dimensional volume rendering of velocity along with
    density iso-surface (wire-mesh). The axises are given in pixels. The
    velocity scale is 1 to 100 km s$^{-1}$ with dark scales representing
    high velocity. The wire mesh shows an
    iso-surface at $\rho = 3.4 \times 10^{-18}$ g cm$^{-3}$.}
  \label{fig:fg2}
\end{figure}
\begin{figure}
  \caption{Cuts in velocity and density. The left top panel is the
    velocity along a vertical line at $x=80$ AU (the center of the AGB star).
    The bottom left panel is the density along the same line.
    The right panels are taken at $x=90$ AU (approximately the
    position of the jet).}
  \label{fig:fg3}
\end{figure}
\begin{figure}
  \caption{Weak Jet greyscale density maps of cross sections in the
    a) $xz$ plane, b) $yz$ plane, and c) $xy$ plane (\ie the
    equatorial plane) with $\chi=1.25$. Panels on both sides of plot
    c) are enlargements of the plots a) and b). The gray scale is the
    logarithm of the density.}
  \label{fig:fg4}
\end{figure}
\begin{figure}
  \caption{Cuts in velocity and density for a $\chi=1.25$
    simulation. The left top panel is the velocity along a vertical
    line at $x=80$ AU ($z$-axis). The bottom left panel is the density
    along the same line . The right panels are similar, but they taken
    are at $x=90$ AU (approximately the position of the jet)}.
  \label{fig:fg5}
\end{figure}
\begin{figure}
  \caption{Greyscale density maps like in the Figure
    \ref{fig:fg1}, but for $\chi=5$.}
  \label{fig:fg6}
\end{figure}
\begin{figure}
  \caption{Cuts in velocity and density as in the Figure
    \ref{fig:fg5} but for $\chi=5.0$}
  \label{fig:fg7}
\end{figure}
\begin{figure}
  \caption{Isosurfaces of the density with values of $6.168 \times
    10^{-18} \rmm{g~cm^{-3}}$, $5.607 \times 10^{-19}\rmm{g~cm^{-3}}$,
    and $1.78 \times 10^{-19} \rmm{g~cm^{-3}}$. The units in the
    axises are given in pixels like in the Figure~\ref{fig:fg2}.}
  \label{fig:fg8}
\end{figure}

\end{document}